\documentclass[sigplan,screen]{acmart}
\usepackage{pifont}
\newcommand{\cmark}{\color{red}{{\small $\checkmark$}}}%
\newcommand{\xmark}{\color{blue}{{\small $\times$}}}%
\AtBeginDocument{%
  \providecommand\BibTeX{{%
    \normalfont B\kern-0.5em{\scshape i\kern-0.25em b}\kern-0.8em\TeX}}}

\copyrightyear{2022}
\acmYear{2022}
\setcopyright{rightsretained}
\acmConference[UIST '22 Adjunct]{The Adjunct Publication of the 35th
Annual ACM Symposium on User Interface Software and
Technology}{October 29-November 2, 2022}{Bend, OR, USA}
\acmBooktitle{The Adjunct Publication of the 35th Annual ACM
Symposium on User Interface Software and Technology (UIST '22 Adjunct),
October 29-November 2, 2022, Bend, OR, USA}
\acmDOI{10.1145/3526114.3558718}
\acmISBN{978-1-4503-9321-8/22/10}

\begin{document}

\title{\textit{DIY Graphics Tab:} A Cost-Effective Alternative to Graphics Tablet for Educators}


\author{Mohammad Imrul Jubair}
\affiliation{%
  \institution{University of Colorado Boulder, USA}
  \city{}
  \country{}
}
\email{mohammad.jubair@colorado.edu}

\author{Arafat Ibne Yousuf}
\affiliation{%
   \institution{Ahsanullah University of Science and Technology, Bangladesh}
  \city{}
  \country{}
 }
 \email{arafat.ysf@gmail.com}

\author{Tashfiq Ahmed}
\affiliation{%
   \institution{Ahsanullah University of Science and Technology, Bangladesh}
  \city{}
  \country{}
  }
  \email{tashfiq.ahm@gmail.com}

\author{Hasanath Jamy}
\affiliation{%
    \institution{Ahsanullah University of Science and Technology, Bangladesh}
  \city{}
  \country{}
  }
\email{jamy.hasanath03@gmail.com}

\author{Foisal Reza}
\affiliation{%
  \institution{Ahsanullah University of Science and Technology, Bangladesh}
  \city{}
  \country{}
 }
\email{foisalreza40@gmail.com}

\author{Mohsena Ashraf}
\affiliation{%
  \institution{University of Colorado Boulder, USA}
  \city{}
  \country{}
 }
\email{mohsena.ashraf@colorado.edu}

\renewcommand{\shortauthors}{Jubair and Yousuf, et al.}

\begin{abstract}
Recording lectures is a normal task for online educators, and a graphics tablet is a great tool for that. However, it is very expensive for many instructors. In this paper, we propose an alternative called ``\textit{DIY Graphics Tab}'' that functions largely in the same way as a graphic tab, but requires only a pen, paper, and laptop's webcam. Our system takes images of writings on a paper with a webcam and outputs the contents. The task is not straightforward since there are obstacles, such as hand occlusion, paper movements, lighting, and perspective distortion due to the viewing angle. A pipeline is used that applies segmentation and post-processing for generating appropriate output from input frames. We also conducted user experience evaluations from the teachers.
\end{abstract}
\vspace{-3mm}
\begin{teaserfigure}
  \includegraphics[width=\textwidth]{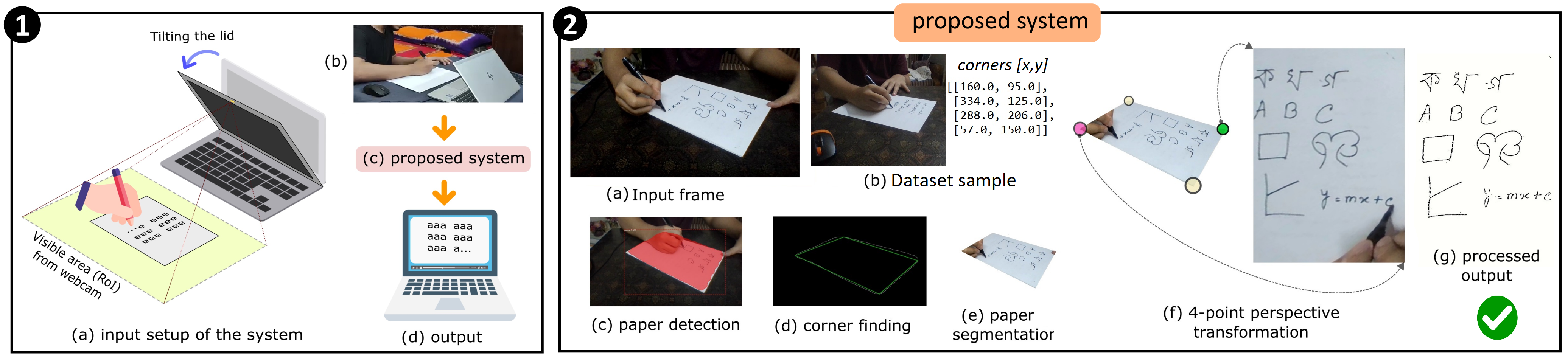}
  \caption{An overview of our \textit{\textbf{DIY Graphics Tab}}. Here, {\protect\includegraphics[scale=.2]{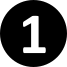}} and \protect\includegraphics[scale=.2]{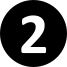} shows the configuration and proposed pipeline respectively.}
  \label{fig:teaser}
\end{teaserfigure}
\maketitle

\section{Introduction}
\label{sec:intro}
In online teaching, it is important for educators to record written contents to make the lectures engaging. Two popular techniques for recording are: \textit{(A)} utilizing an overhead camera which is straightforward, and \textit{(B)} using a graphics tablet or tab~\cite{Tab2021}, i.e. \textit{Intuos}\footnote{\texttt{www.wacom.com/en-us/products/pen-tablets}}.
These tabs are becoming popular, but many teacher find it difficult as they are expensive~\cite{linda_2018}, and many
lack abilities to use such instruments. Besides, most general-purpose graphics tabs have no built-in screen; hence, users must maintain 
\textit{hand-eye synchronization}~\cite{santos_2020}, which can be displeasing. Excessive use of these gadgets might lead to illness, like \textit{musculoskeletal}.
~\cite{Xu2020}. Tablets with integrated touch displays are even more expensive.

In this paper, we present an alternative to graphics tab which is affordable yet providing its essential features. Our goal is to combine the old-fashioned pen-paper technique with the modern computer-based recording technology.
Our solution avoids huge budget, and does not necessitate hand-eye synchronization. It needs only a \textbf{laptop}, a \textbf{pen}, and \textbf{paper}; and we name it ``\textit{{Do-It-Yourself Graphics Tab}}'', or ``\textit{{DIY Graphics Tab}}''.
Fig.~\ref{fig:teaser} presents an overview \textit{DIY Graphics Tab}. To configure, user places a paper in front of a laptop and tilts its lid so that the webcam can capture the area containing
the paper $\bigl[$\protect\includegraphics[scale=.19]{crc1.png}{$^{\large (a, b)}\bigr]$}. We found that, a tilting angle around $45$ degrees \textit{w.r.t} the base is adequate.
Our system then processes the frame, and only contents are rendered on screen---as if it were taken from a ``bird's eye" view $\bigl[$\protect\includegraphics[scale=.19]{crc1.png}{$^{\large (d)}\bigr]$}. Our system takes care of the obstacles, such as---occlusion by hand, random movement of paper, background, lighting conditions, and perspective distortion due to the angular view.\\

Few works were done on capturing lectures and analyzing the contents, i.e.,
\cite{Davila2021, UralaKota2019, kota2018automated, davila2017whiteboard, lee2017robust, Yeh2014, Wienecke2003}, but they focused on whiteboard. Research that inclines mostly with our domain is \textit{WebcamPaperPen}~\cite{webcamPP2014}. In terms of ease of use and simplicity in configuration, our work outperforms \textit{WPP} (Table~\ref{tab:compare}).
\begin{table}[h]
\centering
\begin{tabular}{lll}
\textit{\textbf{\small Negative features}} &
  \textit{\textbf{\small WPP}} &
  \textit{\textbf{\small DIYGT}} \\ \hline \vspace{-1mm}
\begin{tabular}[c]{@{}l@{}}\small Requires manual detection? \end{tabular} &
  {\cmark} &
  \begin{tabular}[c]{@{}l@{}}{\xmark}\end{tabular} \\ \vspace{-1mm}
\begin{tabular}[c]{@{}l@{}}\small Requires extra equipment?\end{tabular} &
  \begin{tabular}[c]{@{}l@{}}{\cmark}\end{tabular} &
  \xmark \\ \vspace{-1mm}
\begin{tabular}[c]{@{}l@{}}\small Requires steadiness?\end{tabular} &
 \cmark &
  \xmark \\ \vspace{-1mm}
\begin{tabular}[c]{@{}l@{}}\small Requires specific type of pen?\end{tabular} &
  \begin{tabular}[c]{@{}l@{}}{\cmark}\end{tabular} &
  \xmark \\ \vspace{-1mm}
\small Restriction on handedness? &
  \begin{tabular}[c]{@{}l@{}}{\cmark}\end{tabular} &
  \xmark \\ \vspace{-1mm}
  \small Requires eye-hand sync? &
  \begin{tabular}[c]{@{}l@{}}{\cmark}\end{tabular} &
  \begin{tabular}[c]{@{}l@{}} {\xmark}\end{tabular} \\
\begin{tabular}[c]{@{}l@{}}\small Requires prior experience?\end{tabular} &
  \cmark &
  \xmark \\ \hline
\end{tabular}
\caption{A comparison between \textit{WPP}~\cite{webcamPP2014} and ours.}
\vspace{-11mm}
\label{tab:compare}
\end{table}
\section{Implementation}
Fig.~\ref{fig:teaser}\protect\includegraphics[scale=.19]{crc2.png} shows our pipeline. After a frame is captured $\bigl[$\protect\includegraphics[scale=.19]{crc2.png}{$^{\large (a)}\bigr]$}, it goes through several stages, which are explained below.

$\bigl[$\protect\includegraphics[scale=.19]{crc2.png}{$^{\large (b, c)}\bigr]$}\textbf{:} To extract paper region, we developed a dataset of images of papers in different positions and lighting, and occlusions. Coordinates of a paper corners are stored as a convex hull of a quadrilateral mask. We trained \textit{Mask-RCNN}~\cite{he2017mask} model using our dataset and performed segmentation.

$\bigl[$\protect\includegraphics[scale=.19]{crc2.png}{$^{\large (d, e)}\bigr]$}\textbf{:} Segmentation provided a mask covering paper. We extracted the largest contour of mask to detect corners.

$\bigl[$\protect\includegraphics[scale=.19]{crc2.png}{$^{\large (f)}\bigr]$}\textbf{:} The \textit{$4$-point perspective transformation} \cite{rosebrock_2014, articleIPM, DBLP:journals/corr/abs-1812-00913} is applied on the corners to unwarp the paper region.

$\bigl[$\protect\includegraphics[scale=.19]{crc2.png}{$^{\large (g)}\bigr]$}\textbf{:} We applied adaptive thresholding~\cite{nina2011recursive} followed by morphological operations and connected component analysis~\cite{dougherty2003hands} to avoid palm, fingers and noises.\\

\begin{figure}[t]
\vspace{-4mm}
  \includegraphics[width=0.4\textwidth]{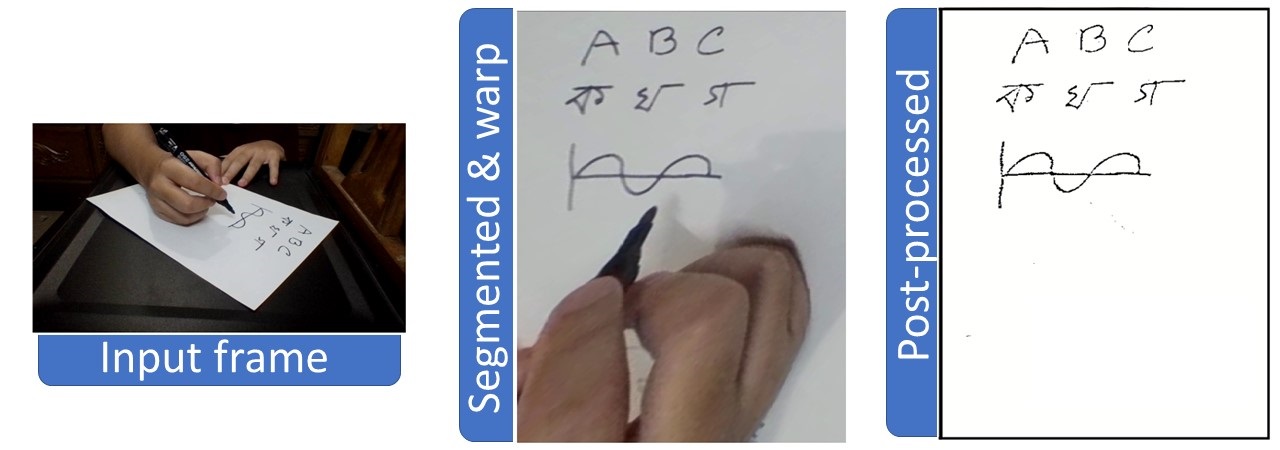}
  \caption{Result of \textit{DIY Graphics Tab}.}
  \label{fig:res}
  \vspace{-4mm}
\end{figure}

To make the system robust, user needs to specify his/her handedness. For a left-handed person our system simply flips the frame in segmentation phase along the axis. Fig.~\ref{fig:res} shows more result. Materials and demos are available \href{https://imruljubair.github.io/project/project-page.html#tab}{{\color{blue}here}}. 
\section{Evaluation}
We performed a survey on $29$ educators from university level, inviting them to use \textit{DIY Graphics Tab} for recording their lectures. We divide them into $3$ groups depending on their technical backgrounds on computer,
which are---\textit{ {\textbf{\color{red}fTB}:}} teachers with no or very few technical background.
\textit{\textbf{\color{blue}TB$-$GT}:} Teachers with technical background with no previous experience of using graphics tablet.
\textit{\textbf{\color{gray}TB$+$GT}:} Teachers with technical background and also has previous experience of using graphics tablet.
We asked participants to utilize our method and grade it on a scale of \textcircled{\small 1} to \textcircled{\small 5} (\textit{very poor} to \textit{excellent})~\cite{likert1932technique, mcleod_1970}.
We found $44.4\%$ of \textit{\textbf{\color{gray}TB$+$GT}} gave \textcircled{\small 4}. Our method received \textcircled{\small 5} 
from a majority ($57.1\%$) of \textit{\textbf{\color{blue}TB$-$GT}}.
Almost $77\%$ of our target users, \textit{\textbf{\color{red}fTB}}, gave highest score. The average rating is $4.44$.

The lid must be slanted in our system which rise concerns as the screen becomes non-visible to instructors. But we are motivated from overhead camera configuration---where the user primarily concentrates on desk. However, we conducted a poll to assess if the tilting may be a problem. We observe that around $72\%$ of teachers are comfortable with it. They think the non-visible screen is not a major concern as long as they can save the costs of graphic tabs.

We also collected a questionnaire-based evaluation from testers to determine the cost-effectiveness of our \textit{DIY Graphics Tab}. To maintain our evaluation on the same ground, we mostly use the questions from \textit{WebcamPaperPen}~\cite{webcamPP2014, mastersthesis}. The user responses are listed below. We observed a majority of users believed that using a graphical tab is highly expensive and found our technology to be cost efficient.
\begin{itemize}
\item \small{It replaces graphics tab perfectly for lecture recording ($31.03\%$)}.
\item \small{The actual graphics tab is better but it is too expensive, I would rather use it ($41.38\%$).}
\item \small{After a certain period I would get tired of it and would buy a proper graphics tab to improve my content ($20.69\%$).}
\item \small{It does not fit my lecture recording ($6.9\%$).}
\end{itemize}
\section{Conclusion}
In this paper, we introduced \textit{DIY Graphics Tab} as a substitute for graphic tab using pen--paper and webcam. We used \textit{Mask-RCNN} to obtain the region of paper image and applied perspective transformation to achieve top-down view.
Our system has shortcomings that demand futher research, e.g., flickering, haziness. etc. We plan to enrich dataset and use advanced segmentation techniques, e.g., \textit{YOLACT}~\cite{bolya2019yolact, yolactplus}. We also like to convert our system into a real-time application.
\newpage
\newpage
\bibliographystyle{ACM-Reference-Format}
\bibliography{sample-base}

\appendix

\end{document}